\documentclass{amsart}
\usepackage{amsfonts}
\usepackage{amsmath,amscd}
\usepackage{amsthm}
\usepackage{amssymb}
\usepackage{latexsym}

\usepackage{color}

\newtheorem{prop}{Proposition}[section]

\newtheorem{example}{Example}
\newtheorem{defn}{Definition}[section]

\newtheorem{rem}{Remark}

\numberwithin{equation}{section}
\begin{document}
\newcommand{\beqa}{\begin{eqnarray}}
\newcommand{\eeqa}{\end{eqnarray}}
\newcommand{\thmref}[1]{Theorem~\ref{#1}}
\newcommand{\secref}[1]{Sect.~\ref{#1}}
\newcommand{\lemref}[1]{Lemma~\ref{#1}}
\newcommand{\propref}[1]{Proposition~\ref{#1}}
\newcommand{\corref}[1]{Corollary~\ref{#1}}
\newcommand{\remref}[1]{Remark~\ref{#1}}
\newcommand{\er}[1]{(\ref{#1})}
\newcommand{\nc}{\newcommand}
\newcommand{\rnc}{\renewcommand}

\nc{\cal}{\mathcal}

\nc{\goth}{\mathfrak}
\rnc{\bold}{\mathbf}
\renewcommand{\frak}{\mathfrak}
\renewcommand{\Bbb}{\mathbb}

\newcommand{\id}{\text{id}}
\nc{\Cal}{\mathcal}
\nc{\Xp}[1]{X^+(#1)}
\nc{\Xm}[1]{X^-(#1)}
\nc{\on}{\operatorname}
\nc{\ch}{\mbox{ch}}
\nc{\Z}{{\bold Z}}
\nc{\J}{{\mathcal J}}
\nc{\C}{{\bold C}}
\nc{\Q}{{\bold Q}}
\nc{\oC}{{\widetilde{C}}}
\nc{\oc}{{\tilde{c}}}
\nc{\og}{{\tilde{\gamma}}}
\nc{\lC}{{\overline{C}}}
\nc{\lc}{{\overline{c}}}
\nc{\Rt}{{\tilde{R}}}

\nc{\odel}{{\overline{\delta}}}

\renewcommand{\P}{{\mathcal P}}
\nc{\N}{{\Bbb N}}
\nc\beq{\begin{equation}}
\nc\enq{\end{equation}}
\nc\lan{\langle}
\nc\ran{\rangle}
\nc\bsl{\backslash}
\nc\mto{\mapsto}
\nc\lra{\leftrightarrow}
\nc\hra{\hookrightarrow}
\nc\sm{\smallmatrix}
\nc\esm{\endsmallmatrix}
\nc\sub{\subset}
\nc\ti{\tilde}
\nc\nl{\newline}
\nc\fra{\frac}
\nc\und{\underline}
\nc\ov{\overline}
\nc\ot{\otimes}
\nc\bbq{\bar{\bq}_l}
\nc\bcc{\thickfracwithdelims[]\thickness0}
\nc\ad{\text{\rm ad}}
\nc\Ad{\text{\rm Ad}}
\nc\Hom{\text{\rm Hom}}
\nc\End{\text{\rm End}}
\nc\Ind{\text{\rm Ind}}
\nc\Res{\text{\rm Res}}
\nc\Ker{\text{\rm Ker}}
\rnc\Im{\text{Im}}
\nc\sgn{\text{\rm sgn}}
\nc\tr{\text{\rm tr}}
\nc\Tr{\text{\rm Tr}}
\nc\supp{\text{\rm supp}}
\nc\card{\text{\rm card}}
\nc\bst{{}^\bigstar\!}
\nc\he{\heartsuit}
\nc\clu{\clubsuit}
\nc\spa{\spadesuit}
\nc\di{\diamond}
\nc\cW{\cal W}
\nc\cG{\cal G}
\nc\al{\alpha}
\nc\bet{\beta}
\nc\ga{\gamma}
\nc\de{\delta}
\nc\ep{\epsilon}
\nc\io{\iota}
\nc\om{\omega}
\nc\si{\sigma}
\rnc\th{\theta}
\nc\ka{\kappa}
\nc\la{\lambda}
\nc\ze{\zeta}

\nc\vp{\varpi}
\nc\vt{\vartheta}
\nc\vr{\varrho}

\nc\Ga{\Gamma}
\nc\De{\Delta}
\nc\Om{\Omega}
\nc\Si{\Sigma}
\nc\Th{\Theta}
\nc\La{\Lambda}

\nc\boa{\bold a}
\nc\bob{\bold b}
\nc\boc{\bold c}
\nc\bod{\bold d}
\nc\boe{\bold e}
\nc\bof{\bold f}
\nc\bog{\bold g}
\nc\boh{\bold h}
\nc\boi{\bold i}
\nc\boj{\bold j}
\nc\bok{\bold k}
\nc\bol{\bold l}
\nc\bom{\bold m}
\nc\bon{\bold n}
\nc\boo{\bold o}
\nc\bop{\bold p}
\nc\boq{\bold q}
\nc\bor{\bold r}
\nc\bos{\bold s}
\nc\bou{\bold u}
\nc\bov{\bold v}
\nc\bow{\bold w}
\nc\boz{\bold z}

\nc\ba{\bold A}
\nc\bb{\bold B}
\nc\bc{\bold C}
\nc\bd{\bold D}
\nc\be{\bold E}
\nc\bg{\bold G}
\nc\bh{\bold H}
\nc\bi{\bold I}
\nc\bj{\bold J}
\nc\bk{\bold K}
\nc\bl{\bold L}
\nc\bm{\bold M}
\nc\bn{\bold N}
\nc\bo{\bold O}
\nc\bp{\bold P}
\nc\bq{\bold Q}
\nc\br{\bold R}
\nc\bs{\bold S}
\nc\bt{\bold T}
\nc\bu{\bold U}
\nc\bv{\bold V}
\nc\bw{\bold W}
\nc\bz{\bold Z}
\nc\bx{\bold X}

\nc\ca{\mathcal A}
\nc\cb{\mathcal B}
\nc\cc{\mathcal C}
\nc\cd{\mathcal D}
\nc\ce{\mathcal E}
\nc\cf{\mathcal F}
\nc\cg{\mathcal G}
\rnc\ch{\mathcal H}
\nc\ci{\mathcal I}
\nc\cj{\mathcal J}
\nc\ck{\mathcal K}
\nc\cl{\mathcal L}
\nc\cm{\mathcal M}
\nc\cn{\mathcal N}
\nc\co{\mathcal O}
\nc\cp{\mathcal P}
\nc\cq{\mathcal Q}
\nc\car{\mathcal R}
\nc\cs{\mathcal S}
\nc\ct{\mathcal T}
\nc\cu{\mathcal U}
\nc\cv{\mathcal V}
\nc\cz{\mathcal Z}
\nc\cx{\mathcal X}
\nc\cy{\mathcal Y}

\nc\e[1]{E_{#1}}
\nc\ei[1]{E_{\delta - \alpha_{#1}}}
\nc\esi[1]{E_{s \delta - \alpha_{#1}}}
\nc\eri[1]{E_{r \delta - \alpha_{#1}}}
\nc\ed[2][]{E_{#1 \delta,#2}}
\nc\ekd[1]{E_{k \delta,#1}}
\nc\emd[1]{E_{m \delta,#1}}
\nc\erd[1]{E_{r \delta,#1}}

\nc\ef[1]{F_{#1}}
\nc\efi[1]{F_{\delta - \alpha_{#1}}}
\nc\efsi[1]{F_{s \delta - \alpha_{#1}}}
\nc\efri[1]{F_{r \delta - \alpha_{#1}}}
\nc\efd[2][]{F_{#1 \delta,#2}}
\nc\efkd[1]{F_{k \delta,#1}}
\nc\efmd[1]{F_{m \delta,#1}}
\nc\efrd[1]{F_{r \delta,#1}}

\nc\fa{\frak a}
\nc\fb{\frak b}
\nc\fc{\frak c}
\nc\fd{\frak d}
\nc\fe{\frak e}
\nc\ff{\frak f}
\nc\fg{\frak g}
\nc\fh{\frak h}
\nc\fj{\frak j}
\nc\fk{\frak k}
\nc\fl{\frak l}
\nc\fm{\frak m}
\nc\fn{\frak n}
\nc\fo{\frak o}
\nc\fp{\frak p}
\nc\fq{\frak q}
\nc\fr{\frak r}
\nc\fs{\frak s}
\nc\ft{\frak t}
\nc\fu{\frak u}
\nc\fv{\frak v}
\nc\fz{\frak z}
\nc\fx{\frak x}
\nc\fy{\frak y}

\nc\fA{\frak A}
\nc\fB{\frak B}
\nc\fC{\frak C}
\nc\fD{\frak D}
\nc\fE{\frak E}
\nc\fF{\frak F}
\nc\fG{\frak G}
\nc\fH{\frak H}
\nc\fJ{\frak J}
\nc\fK{\frak K}
\nc\fL{\frak L}
\nc\fM{\frak M}
\nc\fN{\frak N}
\nc\fO{\frak O}
\nc\fP{\frak P}
\nc\fQ{\frak Q}
\nc\fR{\frak R}
\nc\fS{\frak S}
\nc\fT{\frak T}
\nc\fU{\frak U}
\nc\fV{\frak V}
\nc\fZ{\frak Z}
\nc\fX{\frak X}
\nc\fY{\frak Y}
\nc\tfi{\ti{\Phi}}
\nc\bF{\bold F}
\rnc\bol{\bold 1}

\nc\ua{\bold U_\A}

\nc\qinti[1]{[#1]_i}
\nc\q[1]{[#1]_q}
\nc\xpm[2]{E_{#2 \delta \pm \alpha_#1}}  
\nc\xmp[2]{E_{#2 \delta \mp \alpha_#1}}
\nc\xp[2]{E_{#2 \delta + \alpha_{#1}}}
\nc\xm[2]{E_{#2 \delta - \alpha_{#1}}}
\nc\hik{\ed{k}{i}}
\nc\hjl{\ed{l}{j}}
\nc\qcoeff[3]{\left[ \begin{smallmatrix} {#1}& \\ {#2}& \end{smallmatrix}
\negthickspace \right]_{#3}}
\nc\qi{q}
\nc\qj{q}

\nc\ufdm{{_\ca\bu}_{\rm fd}^{\le 0}}


\nc\isom{\cong} 

\nc{\pone}{{\Bbb C}{\Bbb P}^1}
\nc{\pa}{\partial}
\def\H{\mathcal H}
\def\L{\mathcal L}
\nc{\F}{{\mathcal F}}
\nc{\Sym}{{\goth S}}
\nc{\A}{{\mathcal A}}
\nc{\arr}{\rightarrow}
\nc{\larr}{\longrightarrow}

\nc{\ri}{\rangle}
\nc{\lef}{\langle}
\nc{\W}{{\mathcal W}}
\nc{\uqatwoatone}{{U_{q,1}}(\su)}
\nc{\uqtwo}{U_q(\goth{sl}_2)}
\nc{\dij}{\delta_{ij}}
\nc{\divei}{E_{\alpha_i}^{(n)}}
\nc{\divfi}{F_{\alpha_i}^{(n)}}
\nc{\Lzero}{\Lambda_0}
\nc{\Lone}{\Lambda_1}
\nc{\ve}{\varepsilon}
\nc{\phioneminusi}{\Phi^{(1-i,i)}}
\nc{\phioneminusistar}{\Phi^{* (1-i,i)}}
\nc{\phii}{\Phi^{(i,1-i)}}
\nc{\Li}{\Lambda_i}
\nc{\Loneminusi}{\Lambda_{1-i}}
\nc{\vtimesz}{v_\ve \otimes z^m}

\nc{\asltwo}{\widehat{\goth{sl}_2}}
\nc\ag{\widehat{\goth{g}}}  
\nc\teb{\tilde E_\boc}
\nc\tebp{\tilde E_{\boc'}}

\newcommand{\LR}{\bar{R}}
\newcommand{\eeq}{\end{equation}}
\newcommand{\ben}{\begin{eqnarray}}
\newcommand{\een}{\end{eqnarray}}

\title[Central extension of the reflection equations]{Central extension of the reflection equations \\ and an analog of Miki's formula}
\author{P. Baseilhac}
\address{Laboratoire de Math\'ematiques et Physique Th\'eorique CNRS/UMR 6083,
     F\'ed\'eration Denis Poisson, Universit\'e de Tours, Parc de Grammont, 37200 Tours, FRANCE}
     \email{baseilha@lmpt.univ-tours.fr}

\author{S. Belliard}
\address{Istituto Nazionale di Fisica Nucleare, Sezione di Bologna, Via Irnerio 46,  40126 Bologna,  Italy}
\email{belliard@bo.infn.it}
\begin{abstract} Two different types of centrally extended quantum reflection algebras are introduced. Realizations in terms of the elements of the central extension of the Yang-Baxter algebra are exhibited. A coaction map is identified. For the special case of $U_q(\widehat{sl_2})$, a realization in terms of elements satisfying the Zamolodchikov-Faddeev algebra - a `boundary' analog of Miki's formula - is also proposed, providing a free field realization of $O_q(\widehat{sl_2})$ ($q-$Onsager) currents.     
\end{abstract}

\maketitle

\vskip -0.6cm

{\small MSC:\ 81R50;\ 81R10;\ 81U15.}

{{\small  {\it \bf Keywords}: Reflection equations; Central extension; Zamolodchikov-Faddeev algebra; Integrable models}}

\section{Introduction}
Since the fundamental works of Drinfeld \cite{D1} and Jimbo \cite{J1} who introduced a certain $q-$deformation of the universal envelopping algebra $U(\widehat{g})$ of any Kac-Moody algebra $\widehat{g}$, there has been an increasing interest for the subject of quantum groups and their numerous applications in the context of  quantum integrable systems. Besides the initial construction of Drinfeld-Jimbo, recall that  Faddeev-Reshetikhin-Takhtadjan \cite{FRT} proposed a realization of $U_q(g)$ by means of solutions of the Yang-Baxter equation. Later on, this framework was extended to the affine extension of $g$.  In addition to \cite{D1,J1}, two other realizations of  the quantum affine algebra  $U_q(\widehat{g})$ were given: one in terms of currents in \cite{D2}, and another one proposed by Reshetikhin and Semenov-Tian-Shansky \cite{RS} who found a way to incorporate the central extension in the formalism of \cite{FRT}. Importantly, these results opened the way to construct explicitely infinite dimensional  representations of $U_q(\widehat{g})$ using $q-$bosons by analogy with the classical case - representations of bosonic or fermionic type \cite{FKJ}. For the class of quantum integrable systems with hidden non-Abelian symmetry associated with $U_q(\widehat{g})$,  this latter formulation became very useful for practical purpose :  for instance, $q-$bosons realizations play a key role in the vertex operators' approach to the solution of the XXZ spin chain in the thermodynamic limit \cite{vertex,JKKKM} (see also \cite{FuKo,Ko}), allowing to derive integral representations for correlation functions of local operators of the model.
Also, central extensions of other algebras enjoying analogous RTT structures such as Yangians became of interest \cite{KLP}, as little was known about their nontrivial infinite dimensional representations that are relevant in the free field approach to certain massive integrable quantum field theory\footnote{For instance, the hidden non-Abelian symmetry of the $SU(2)-$invariant Thirring model is the Yangian double \cite{S2}.} \cite {Luk}.\vspace{1mm}

Other types of algebras - tridiagonal algebras and the so-called $q-$Onsager algebras -  have recently attracted attention either in the context of $q-$orthogonal polynomials, generalized special functions and related classification schemes \cite{Ter93,Ter01,Ter03} (see \cite{Ter04} and references therein), or in the context of integrable systems on the lattice or continuum, with or without boundaries \cite{B1,Bas2,BK1,BK2,BasB}. For instance, in the simplest case corresponding to $g=sl_2$, the $q-$Onsager algebra \cite{Ter93} is known to be isomorphic as a coideal algebra \cite{BasS} to the infinite dimensional (current) algebra equiped with a certain coaction map - here denoted  $O_q(\widehat{sl_2})$ -  introduced in \cite{BK} and related reflection equation algebra \cite{Cher84,Skly88} with $U_q(\widehat{sl_2})$ R-matrix.  Besides the interest of such isomorphisms from the mathematical point of view, they provide an efficient way to construct certain representations of the $q-$Onsager algebra  that may be useful in solving quantum integrable systems: explicit  `dressed' solutions of the reflection equations were used to derive finite dimensional tensor product representations of the $q-$Onsager algebra\footnote{For $q$ not a root of unity, finite dimensional representations of the $q-$Onsager algebra have been classified in \cite{Ter04}.} \cite{BK,BK1}. Based on the remarkable properties of these representations (see \cite{Ter03} for details), an exact solution\footnote{Although not explored yet, another type of approach using the $q-$Onsager algebra and related representation theory may be considered following some ideas of Davies \cite{Da}.} for the spectrum and eigenstates of the finite XXZ open spin chain with {\it generic} boundary conditions\footnote{For certain non-diagonal boundary conditions, diagonal ones or $q$ a root of unity, a Bethe ansatz solution exists and has  been considered \cite{Skly88,bXXZ}. For {\it generic} non-diagonal boundary conditions, note the functional approach considered in \cite{Galleas}.} was proposed in \cite{BK2} using the results of \cite{TDpair}. In this approach, the spectral problem for all integrals of motions is reduced to the diagonalisation of a single block tridiagonal matrix without spectral parameter. However, although important simplifications may be considered following \cite{Molinari} for a finite dimensional space of states, applying a similar approach to the semi-infinite XXZ spin chain with generic non-diagonal boundary conditions requires to study suitable infinite dimensional representations of the $q-$Onsager algebra.\vspace{1mm} 

More generally, in order to further develop an approach `{\it \`a la Onsager}' \cite{Bas0} in the spirit of \cite{Ons44,Da,XY,Ahn,Gehlen,Baxter,Gehlen2} that would find applications in integrable non-conformal quantum field theories or in the thermodynamic limit of lattice integrable models enjoying $q-$Onsager's type of non-Abelian symmetry, previous experience suggests to answer the following questions:  what is the central extension of the reflection equation algebra and what kind of nontrivial infinite dimensional representations of associated current algebra may be derived in this framework? For instance, having in mind that the current algebra $O_q(\widehat{sl_2})$ is a certain coideal subalgebra\footnote{For definitions, see \cite{MRS}. Note that the explicit form of the coaction of $O_q(\widehat{sl_2})$ currents is given in \cite{BasS}, extending to the infinite dimensional case the results of \cite{BK1}.} of $U_q(\widehat{sl_2})$,  by analogy with the finite dimensional case there are good reasons to expect that  introducing a central extension of the reflection equation algebra and constructing solutions would provide efficient tools to derive nontrivial infinite dimensional representations of  $O_q(\widehat{sl_2})$ with nonzero central charge. Indeed, in the case of $U_q(\widehat{sl_2})$ recall that $L-$operators satisfying the extended Yang-Baxter algebra \cite{RS} can be actually realized in terms of type I and II vertex operators thanks to a formula proposed by Miki \cite{Miki} - see also \cite{Pak}. In the context of quantum integrable models, considering a central extension of the reflection equation algebra - a problem already suggested fifteen years ago in \cite{JKKKM} -  and building Miki's type of solutions would open the possibility to derive bosonic type of representations that would find straightforward applications in studying the thermodynamic limit of boundary lattice integrable models with $q-$Onsager non-Abelian symmetry. Motivated by these open problems, some answers to the above-mentionned questions are proposed in the present letter.
\vspace{1mm} 

Here, our purpose is twofold. First, by analogy with Reshetikhin-Semenov-Tian-Shansky \cite{RS} we incorporate a central extension in Cherednik-Sklyanin's formalism \cite{Cher84,Skly88}. Contrary to previous works\footnote{Certain quadratic combinations of $L_{\pm}-$operators were shown to satisfy reflection equations' type of algebra, but the central charge didn't arise explicitly in the algebra's defining relations \cite{MRS}.}, the parameter associated with the nonzero central charge explicitely appears in the defining relations of the new algebra. A coaction map is then identified, which allows to build up `dressed' solutions of the centrally extended reflection equations. 
Secondly, in Section 3 we focus on the special case associated with $U_q(\widehat{sl_2})$ $R-$matrix and nonzero central charge. An analog of Miki's formula is proposed: elements of the extended reflection equation algebra are realized in terms of a single 
one-parameter family of solutions to the reflection equation algebra. Using the results of \cite{BasS}, a realization of $O_q(\widehat{sl_2})$ currents in terms of Zamolodchikov-Faddeev operators follows, which provides a bosonization scheme for the $q-$Onsager algebra.
Comments and further applications of the results are briefly presented in the last Section.\vspace{1mm}
 
\section{Central extension of reflection equation algebras}
The purpose of this Section is to propose a central extension of two different types of quantum reflection equation algebras, coming along with the extension discussed by Reshetikhin and Semenov-Tian-Shansky \cite{RS} in the case of the Yang-Baxter algebra and quantum affine algebras. First, we recall basic definitions that will be used below.\vspace{1mm} 
  
Let the operator-valued function $R:{\mathbb C}^*\mapsto \mathrm{End}({\cal V}\otimes {\cal V})$ be  the intertwining operator (quantum $R-$matrix) between the tensor product of two finite dimensional representation space ${\cal V}$ associated with the algebra $U_q(\widehat{g})$ \cite{D1,J1}. The element $R(u)$ depends on the deformation parameter $q$ and $u$ is called the spectral parameter. Then $R(u)$ satisfies the quantum Yang-Baxter equation in the space $\mathrm{End}({\cal V}_1\otimes {\cal V}_2\otimes {\cal V}_3)$. Using the standard notation $R_{ij}(u)\in \mathrm{End}({\cal V}_i\otimes {\cal V}_j)$, it reads 
\begin{align}
R_{12}(u/v)R_{13}(u)R_{23}(v)=R_{23}(v)R_{13}(u)R_{12}(u/v)\ \qquad \forall u,v\ .\label{YB}
\end{align}
Explicit expressions for $R(u)$ can be found in \cite{J1}. Other examples can be found in the litterature on the subject, where systematic procedures based on the quantum group structure associated with (\ref{YB}) have been thoroughly used  to derive new solutions $R(u)$. See e.g. \cite{KKRS,DGZ}.\vspace{1mm}

Given a solution $R(u)$ of (\ref{YB}), the so-called $L-$operators satisfy the Yang-Baxter algebra $A_q(R)$ \cite{STF,FRT}
\begin{align}
R_{12}(u/v)L_{1}(u)L_{2}(v)=L_{2}(v)L_{1}(u)R_{12}(u/v)\ \qquad \forall u,v\ ,\label{YBalg}
\end{align}
which is known to play a crucial role in the theory of quantum groups and quantum integrable models. A
central extension which defining relations generalize the ones for the finite dimensional quantum enveloping Lie algebras \cite{FRT} has been proposed in \cite{RS}:
\begin{defn}[Central extension of the Yang-Baxter algebra \cite{RS}]
${\widetilde A}_q(R)$ is the associative algebra with unit $1$ and  elements  $L^\pm(u)=(L^\pm_{ij}(u))$
satisfying the defining relations:
\ben
\tilde{R}_{12}(u/v)L_1^\pm(u)L_2^\pm(v)&=&L_2^\pm(v)L_1^\pm(u)\tilde{R}_{12}(u/v)\ ,\nonumber\\
\tilde{R}_{12}(\gamma^{\pm2}u/v)L_1^\mp(u)L_2^\pm(v)&=&L_2^\pm(v)L_1^\mp(u)\tilde{R}_{12}(\gamma^{\mp2}u/v)\label{RLL}
\een
where $\gamma=q^{-c/2}$ and $c$ is called the central extension.  
\end{defn}
\begin{rem}
The $R-$matrix $\tilde{R}(u)$ is obtained from the universal $R-$matrix \cite{FR} by specializing the representation space ${\cal V}$.
For a two-dimensional space and $U_q(\widehat{sl_2})$, it reads as a product of a matrix that satisfies the Yang-Baxter equation (\ref{YB}) times a scalar factor, which is such that 
 $\tilde{R}(u)$  satisfy the relations (\ref{unit}) and (\ref{wcross}).
\end{rem}
Note that the algebra ${\widetilde A}_q(R)$ is equiped with the structure of a Hopf algebra. The coproduct\footnote{The space ordering is changed compared with \cite{RS}.} $\Delta$ is such that:
\ben
\Delta(L^\pm(u))=L^\pm(u \ 1 \otimes \gamma^{\mp 1})\dot{\otimes}L^\pm(u \ \gamma^{\pm 1} \otimes 1)\ , \qquad  \Delta(\gamma)=\gamma \otimes \gamma \ , \nonumber 
\een
 the antipode $S$ and the co-unit $\cal E$ act as:
\ben
S(L^\pm(u))= L^\pm(u)^{-1}\ ,  \quad \cal E(L^\pm(u))=1\ . \nonumber
\een

Other families of quadratic algebras have been introduced  in the context of quantum integrable systems with boundaries  \cite{Cher84,Skly88}. For instance, given a solution of the quantum Yang-Baxter equation (\ref{YB}) the quantum reflection equation algebra
 ${B}_q(R)$
\begin{align} R_{12}(u/v)\ K_1(u)\ R_{21}(uv)\ K_2(v)\
= \  K_2(v)\ R_{12}(uv)\ K_1(u)\ R_{21}(u/v)\  \qquad \forall u,v\ 
\label{REsimple} \end{align}
has been introduced. Also, another family of quantum reflection equation algebra  \cite{MRS,GM} - here denoted ${TB}_q(R)$ -
has been proposed, extending algebraic structures previously considered \cite{Ols}\footnote{See related works \cite{Mo,Mu}.}: 
\begin{align} R_{12}(u/v)\ K_1(u)\ R^{t_1}_{12}(1/uv)\ K_2(v)\
= \  K_2(v)\ R^{t_1}_{12}(1/uv) \ K_1(u)\ R_{12}(u/v)\  \qquad \forall u,v\  .
\label{REsimple2} \end{align}
For both families, explicit solutions $K(u)$ are known  in many cases (see for example \cite{GZ,DV,MLS,DelG,Gand}), but a systematic derivation by analogy with \cite{J1,DGZ} requires to identify explicitely the underlying elementary quantum algebra - without spectral parameter - associated with (\ref{REsimple}) or (\ref{REsimple2}) and study it in details. In the simplest case of $U_q(\widehat{sl_2})$ $R-$matrix, according to a certain choice of coaction map such relations were identified in \cite{B1,Bas2} and found to coincide with a special case of the tridiagonal algebra introduced by Terwilliger in \cite{Ter93} (see related works \cite{Ter01,Ter03,Ter04}) - called the $q-$Onsager algebra.
 For other affine Lie algebras, note that analogous structures have been exhibited in \cite{BasB}. 

\vspace{1mm} 

By analogy with \cite{RS}, it is thus natural to introduce the following central extension of the algebra (\ref{REsimple}):
\begin{defn}[Central extension of the reflection equation algebra (\ref{REsimple})] ${\widetilde B}_q(R)$ is the associative algebra with unit $1$ and elements 
$K^{(\epsilon_1\epsilon_2)}(u)=(K^{(\epsilon_1\epsilon_2)}_{ij}(u))$ 
 satisfying the defining relations:
\ben
\tilde{R}_{12}(u/v)\ K^{(++)}_1(u)\ \tilde{R}_{21}(uv)\  K^{(++)}_2(v)\
&=& \ K^{(++)}_2(v)\ \tilde{R}_{12}(uv)\ K^{(++)}_1(u)\ \tilde{R}_{21}(u/v)\ ,\nonumber\\
\tilde{R}_{12}(\og^{-2}u/v)\ K^{(++)}_1(u)\ \tilde{R}_{21}(\og^{2}uv)\  K^{(+-)}_2(v)\
&=& \ K^{(+-)}_2(v)\ R_{12}(\og^{-2}uv)\ K^{(++)}_1(u)\ R_{21}(\og^{2}u/v)\ ,\nonumber\\
\tilde{R}_{12}(\og^{-2}u/v)\ K^{(++)}_1(u)\ \tilde{R}_{21}(\og^{2}uv)\  K^{(-+)}_2(v)\
&=& \ K^{(-+)}_2(v)\ \tilde{R}_{12}(\og^{-2}uv)\ K^{(++)}_1(u)\ \tilde{R}_{21}(\og^{2}u/v)\ ,\nonumber\\
\tilde{R}_{12}(\og^{-4}u/v)\ K^{(++)}_1(u)\ R_{12}(\og^{4}uv)\  K^{(--)}_2(v)\
&=& \ K^{(--)}_2(v)\ \tilde{R}_{12}(\og^{-4}uv)\ K^{(++)}_1(u)\ \tilde{R}_{21}(\og^{4}u/v)\ ,\nonumber
\een
\ben
\tilde{R}_{12}(\og^{2}u/v)\ K^{(+-)}_1(u)\ \tilde{R}_{21}(\og^{-2}uv)\  K^{(++)}_2(v)\
&=& \ K^{(++)}_2(v)\ \tilde{R}_{12}(\og^{2}uv)\ K^{(+-)}_1(u)\ \tilde{R}_{21}(\og^{-2}u/v)\ ,\nonumber\\
\tilde{R}_{12}(u/v)\ K^{(+-)}_1(u)\ \tilde{R}_{21}(uv)\  K^{(+-)}_2(v)\
&=& \ K^{(+-)}_2(v)\ \tilde{R}_{12}(uv)\ K^{(+-)}_1(u)\ \tilde{R}_{21}(u/v)\ ,\nonumber\\
\tilde{R}_{12}(u/v)\ K^{(+-)}_1(u)\ \tilde{R}_{21}(uv)\  K^{(-+)}_2(v)\
&=& \ K^{(-+)}_2(v)\ \tilde{R}_{12}(uv)\ K^{(+-)}_1(u)\ \tilde{R}_{21}(u/v)\ ,\nonumber\\
\tilde{R}_{12}(\og^{-2}u/v)\ K^{(+-)}_1(u)\ \tilde{R}_{21}(\og^{2}uv)\  K^{(--)}_2(v)\
&=& \ K^{(--)}_2(v)\ \tilde{R}_{12}(\og^{-2}uv)\ K^{(+-)}_1(u)\ \tilde{R}_{21}(\og^{2}u/v)\ \nonumber
\een
where $\tilde{\gamma}=q^{-\oc/2}$ and $\oc$ is called the central extension.
Defining relations for $K^{(--)}(u)$ with $K^{(++)}(u)$, $K^{(+-)}(u)$, $K^{(-+)}(u)$, $K^{(--)}(u)$  (respectively, for $K^{(-+)}(u)$ with $K^{(++)}(u)$, $K^{(+-)}(u)$, $K^{(-+)}(u)$, $K^{(--)}(u)$) are obtained 
by substituing $+ \leftrightarrow -$ and $\og \rightarrow \og^{-1}$ in the first (second) set of four equations. 
\end{defn}
\begin{defn}[Central extension of the reflection equation algebra  (\ref{REsimple2})]
Defining relations of the central extension $\widetilde{TB}_q(R)$ of the reflection equation algebra ${TB}_q(R)$ (\ref{REsimple2}) \cite{MRS} are introduced analogously: one changes first $R_{21} \to R_{12}$, and then $R_{12}(a uv) \to R^{t_1}_{12}(\frac{1}{a uv})$ in the definition $2.2$. 
\end{defn}
\begin{rem}
If a crossing relation relating the R-matrix  and transposed R-matrix of the form  (\ref{scross}) is satisfied, then the defining relations of the two algebras $\widetilde{B}_q(R)$ and $\widetilde{TB}_q(R)$ become identical. This situation happens for $U_q(\widehat{sl_2})$, the case considered in Section 3.
\end{rem}

Solutions to the equations above can be easily constructed combining the results of \cite{RS} together with \cite{Skly88}, for instance starting from scalar solutions of the reflection equations (\ref{REsimple}) as considered in \cite{GZ,DV,MLS,DelG,Gand} using the so-called `dressing' procedure. 
\begin{example} Let $L^\pm(u)$ be elements of ${\widetilde A}_q(R)$.
 Let $K^{(0)}(u)$ be an element of the subalgebra $B_q(R)$ (\ref{REsimple})  of  ${\widetilde B}_q(R)$.
Then the dressed elements:  
\ben
K^{(\pm\pm)}(u)&\equiv&L^\pm(u\gamma^{\mp 1}) \dot{\otimes} K^{(0)} (u)\dot{\otimes}S(L^{\pm}(u^{-1}\gamma^{\mp 1})) \ ,\label{Ktwist1}\\
K^{(\pm\mp)}(u)&\equiv&L^\pm(u\gamma^{\pm 1})\dot{\otimes}  K^{(0)}(\gamma^{\pm 2}u)\dot{\otimes}S(L^{\mp}(u^{-1}\gamma^{\mp 1}))\label{Ktwist2}
\een
are in ${\widetilde A}_q(R)\otimes {B}_q(R)$ and satisfy the relations of $\widetilde{B}_q(R)$ with the identification $\og=\gamma$. 
\end{example}
\begin{proof} Recall that $L^\pm(u)$ are invertible and $S(L^\pm(u))=L^\pm(u)^{-1}$ \cite{RS}. Then, straightforward calculations using (\ref{RLL}) show the assumption.
\end{proof}

\begin{rem} Combinations of $L-$operators have been previously considered in the litterature. For instance, in \cite{RS} a `quantum current' $L(u)=L^+(u\gamma^{-2}) S(L^-(u))$ has been introduced. It satisfies the relation :
\ben
R(u/v)L_1(u)R_{21}(v/u \gamma^{-4})L_2(v)=L_2(v)R_{21}(v/u \gamma^{4})L_1(u)R(u/v)\ .\label{QCRE}
\een
Note that this quantum version of the Schwinger commutation relation for current algebra on the line essentially differs (see arguments in $R-$matrices) from reflection's type of equations. 
\end{rem}

Quantum affine algebras are known to be Hopf algebras, thanks to the existence of a coproduct, counit and antipode actions. 
For reflection equation's type of algebra,  an analog of Hopf's algebra coproduct called coaction map \cite{Chari} can be easily identified: 
\begin{prop} Let $L^\pm(u)$ be elements of ${\widetilde A}_q(R)$. 
The homomorphism
$\delta: {\widetilde B}_q(R) \rightarrow {\widetilde A}_q(R) \otimes {\widetilde B}_q(R)$
defines a left coaction map of ${\widetilde B}_q(R)$: 
\ben
\delta(\og)&=& \gamma  \otimes \og \ ,\\
\delta\Big({K^{(\pm\pm)}(u)}\Big)&=&L^\pm(u\,\gamma^{\mp 1} \otimes \og^{\mp 2})\dot{\otimes}K^{(\pm\pm)}(u)\dot{\otimes} S(L^{\pm}(u^{-1}\,\gamma^{\mp 1} \otimes  \og^{\mp 2})) \ ,\label{Kd1}\\
\delta \Big({K^{(\pm\mp)}(u)}\Big)&=&L^\pm(u\ \gamma^{\pm 1} \otimes 1)\dot{\otimes}K^{(\pm\mp)}(u\gamma^{\mp 2} \otimes 1)\dot{\otimes}S(L^{\mp}(u^{-1}\,  \gamma^{\mp 1} \otimes 1))\ .
\label{Kd2}
\een
\end{prop}
\begin{prop}
For the central extension $\widetilde{TB}_q(R)$, dressed elements and a coaction map are obtained from Example $1$ and Proposition $2.1$ changing $S(L^\pm(u))\to (L^\pm(u))^t$ in each expression.
\end{prop}

\begin{rem}
Following \cite{DF}, if one assumes $L^\pm(u)=\sum_{n=0}^\infty L^\pm[n]  u^{\pm n}$ in (\ref{Ktwist1}), (\ref{Ktwist2}) (see also examples in \cite{RS}), the elements $K^{(\pm\pm)}(u)$ are currents and
$K^{(\mp\pm)}(u)$ half-currents. The algebras $\widetilde{B}_q(R)$ and $\widetilde{TB}_q(R)$ must be understood as some extended algebra of formal series, similarly to (\ref{QCRE}).
\end{rem}
\begin{rem}
Note that the subalgebras of $\widetilde{B}_q(R)$ (or $\widetilde{TB}_q(R)$) generated by the elements ${K^{(\pm\mp)}(u)}$ have been introduced in \cite{ACDFR} and \cite{MRS}, respectively.
\end{rem}

According to these results, elements of the new algebra can be realized in terms of Ding-Frenkel currents using the explicit form of $L_\pm(u)$ \cite{DF}. For practical purpose,  this `dressing' procedure \cite{Skly88} allows to derive realizations acting on finite dimensional tensor product  representations  (see e.g. \cite{BK,BK1}) starting from a scalar solution. Instead of considering this option, below we will focus on another type of realization that is associated with infinite dimensional representations.

\section{An analog of Miki's formula and $O_q(\widehat{sl_2})$ currents}
In the special case of $U_q(\widehat{sl_2})$ and $c=1$,  in \cite{Miki} Miki showed that $L-$operators satisfying (\ref{RLL}) can be expressed in terms of elements of the Zamolodchikov-Faddeev algebra associated with the quantum $R-$matrix acting on the tensor product of two finite dimensional representations (see also \cite{Pak} for details).  Actually, it is possible to exhibit an analogous realization of $K-$operators that are elements of the central extension of the reflection equation algebra given by Definition 2.2. 
Let $\bar{R}(u)$ be the scalar $R-$matrix which satisfies the crossing symmetry (\ref{scross}) as defined in Appendix A.

\begin{defn} The Zamolodchikov-Faddeev algebra is generated by the operators $\Phi_\pm(u)$ that satisfy
\ben
&&\Phi_2(v)\Phi_1(u)=\bar{R}_{12}(u/v)\Phi_1(u)\Phi_2(v) \quad with \quad\Phi(u)=\left(\begin{array}{c} \Phi_+(u)  \\ \Phi_-(u) \end{array} \right)\ .\label{FZ}
\een
\end{defn}

In Miki's work \cite{Miki}, entries of $L_\pm-$operators were written as certain quadratic combinations mixing elements of two copies of the Zamolodchikov-Faddeev algebra, with simple exchange relations between the elements of the two copies. Here, the situation is even simpler:\vspace{1mm}  
\begin{prop} 
Let $a,w \in {\mathbb C}^*$. Let $\Phi_\pm(u)$ be the components of the Zamolochikov-Faddeev operators and $M$ the matrix (\ref{scross}). Define:
\ben
K(u;a)=\left(\begin{array}{lcl} \Phi_+(ua)\Phi_+(u^{-1}q^{-1}a) &  \Phi_+(ua)\Phi_-(u^{-1}q^{-1}a)  \\ \Phi_-(ua)\Phi_+(u^{-1}q^{-1}a) &  \Phi_-(ua)\Phi_-(u^{-1}q^{-1}a)  \end{array} \right)M\label{Ku} \ .
\een 
Then, $K(u;a_{\epsilon_1\epsilon_2})$ gives a realization of ${\widetilde B}_q(R)$  provided one identifies $K^{(\epsilon_1\epsilon_2)}(u)\equiv K(u;a_{\epsilon_1\epsilon_2})$ with $a_{\pm\pm}={\tilde\gamma}^{\pm 2}w,a_{\pm\mp}=w$ and $\tilde{\gamma}^2=q^k$ with $k \in {\mathbb Z}$.  
\end{prop}
\begin{proof}
Plug (\ref{Ku}) in the defining relations of Definition 2.2 and simplify using the commutation relations (\ref{FZ}). According to the definitions of $\tilde{R}(u)$ and $\bar{R}(u)$ given in the Appendix, together with eq. (\ref{ratios}), the quantization condition $\tilde{c}\equiv k$ follows. 
\end{proof}
\begin{rem}
A generating function for mutually commuting quantities can be constructed using (\ref{Ku}). Note the closed analogy with the  renormalized' transfer matrix of the semi-infinite XXZ open spin chain conjectured in \cite{JKKKM}, in which case Zamolodchikov-Faddeev operators are realized as type I vertex operators of $U_q(\widehat{sl_2})$ at level one ($\tilde{c}=c = 1$). 
\end{rem}

The boundary analog of Miki's formula (\ref{Ku}) here proposed can be constructed either using the properties of intertwiners of the current algebra $O_q(\widehat{sl_2})$  (see details elsewhere \cite{BasB2}) or, alternatively, using original Miki's formula for $L^{\pm}(u)$ in the dressed elements (\ref{Ktwist1}), (\ref{Ktwist2}) with a scalar matrix $K^{(0)}(u)$. In this later case, one of the two copies of the Zamolodchikov-Faddeev algebras generates central elements.\vspace{1mm}

Finally, note that the full set of defining relations determines uniquely the ratios of $`a'-$parameters and the central charge $\tilde{c}$ to take discrete values. However, restricting the analysis to the subalgebra  (\ref{REsimple}) with $K(u)\equiv K(u;a)$, these conditions are no longer required. In this case, $K(u;a)$ provides a one-parameter family of solutions to the reflection equations (\ref{REsimple}) for arbitrary $a$. Let us then consider a straightforward application of this result. Recall that the current algebra $O_q(\widehat{sl_2})$  associated with  (\ref{REsimple}) is given by\footnote{The $q-$commutator $\big[X,Y\big]_q=qXY-q^{-1}YX$ is introduced.} \cite{BasS}:
\begin{defn}[Reflection equation current algebra]{\label{defnCA}} $O_q(\widehat{sl_2})$ is an associative algebra with unit $1$, current generators $\cW_\pm(u)$, $\cG_\pm(u)$ and parameter $\rho\in{\mathbb C}^*$. Define the formal variables $U=(qu^2+q^{-1}u^{-2})/(q+q^{-1})$ and $V=(qv^2+q^{-1}v^{-2})/(q+q^{-1})$ \ $\forall u,v$. The defining relations are:
\begin{align}
&&\big[{\cW}_\pm(u),{\cW}_\pm(v)\big]=0\ ,\qquad\qquad\qquad\qquad\qquad\qquad\qquad\label{ec1}\\
&&\big[{\cW}_+(u),{\cW}_-(v)\big]+\big[{\cW}_-(u),{\cW}_+(v)\big]=0\ ,\qquad\qquad\qquad\qquad\qquad\qquad\qquad\label{ec3}\\
&&(U-V)\big[{\cW}_\pm(u),{\cW}_\mp(v)\big]= \frac{(q-q^{-1})}{\rho(q+q^{-1})}\left({\cG}_\pm(u){\cG}_\mp(v)-{\cG}_\pm(v){\cG}_\mp(u)\right)\qquad\qquad\qquad\label{ec4}\\
&& \qquad \qquad\qquad\qquad\qquad\qquad\qquad\qquad+ \frac{1}{(q+q^{-1})} \big({\cG}_\pm(u)-{\cG}_\mp(u)+{\cG}_\mp(v)-{\cG}_\pm(v)\big)\ ,\nonumber
\end{align}
\beqa
&&{\cW}_\pm(u){\cW}_\pm(v)-{\cW}_\mp(u){\cW}_\mp(v)+\frac{1}{\rho(q^2-q^{-2})}\big[{\cG}_\pm(u),{\cG}_\mp(v)\big]\qquad\qquad\qquad\qquad \qquad\label{ec5}\\
&&\qquad\qquad\qquad\qquad\qquad\qquad+ \ \frac{1-UV}{U-V}\big({\cW}_\pm(u){\cW}_\mp(v)-{\cW}_\pm(v){\cW}_\mp(u)\big)=0\ ,\nonumber\\
&&U\big[{\cG}_\mp(v),{\cW}_\pm(u)\big]_q -V\big[{\cG}_\mp(u),{\cW}_\pm(v)\big]_q - (q-q^{-1})\big({\cW}_\mp(u){\cG}_\mp(v)-{\cW}_\mp(v){\cG}_\mp(u)\big)\label{ec6}\\
&&\qquad\qquad\qquad\qquad\qquad\qquad\qquad \quad + \ \rho \big(U{\cW}_\pm(u)-V{\cW}_\pm(v)-{\cW}_\mp(u)+{\cW}_\mp(v)\big)=0\ ,\nonumber\\
&&U\big[{\cW}_\mp(u),{\cG}_\mp(v)\big]_q -V\big[{\cW}_\mp(v),{\cG}_\mp(u)\big]_q - (q-q^{-1})\big({\cW}_\pm(u){\cG}_\mp(v)-{\cW}_\pm(v){\cG}_\mp(u)\big)\label{ec7}\\
&& \qquad\qquad\qquad\qquad\qquad\qquad\qquad\quad+  \ \rho \big(U{\cW}_\mp(u)-V{\cW}_\mp(v)-{\cW}_\pm(u)+{\cW}_\pm(v)\big)=0\ ,\nonumber\\
&&\big[{\cG}_\epsilon(u),{\cW}_\pm(v)\big]+\big[{\cW}_\pm(u),{\cG}_\epsilon(v)\big]=0 \ ,\quad \forall \epsilon=\pm\label{ec8}\ ,\qquad\qquad\qquad\qquad\qquad\qquad\qquad\\
&&\big[{\cG}_\pm(u),{\cG}_\pm(v)\big]=0\ ,\label{ec9}\qquad\qquad\qquad\qquad\qquad\qquad\qquad\\ 
&&\big[{\cG}_+(u),{\cG}_-(v)\big]+\big[{\cG}_-(u),{\cG}_+(v)\big]=0\ .\qquad\qquad\qquad\qquad\qquad\qquad\qquad\ \label{ec16}
\eeqa
\end{defn}
\vspace{0.2cm}

In \cite{Pak}, realizations of Ding-Frenkel currents were obtained in terms of Zamolodchikov-Faddeev operators. Here, the coideal isomorphism between (\ref{REsimple}) and (\ref{defnCA}) \cite{BasS}  leads to the following result:
\begin{prop} Let $a,k_\pm\in {\mathbb C}^*$ be arbitrary. Setting $\rho\equiv k_+k_-(q+q^{-1})^2$, $O_q(\widehat{sl_2})$  currents admit the one-parameter family of realizations in terms of  the Zamolodchikov-Faddeev operators:
\ben
\cW_+(u)&=& -i \frac{\,q\,u\,\Phi_+(a \, u)\Phi_-(aq^{-1} u^{-1})- q^{-1} u^{-1}\,\Phi_-( a\,u)\Phi_+(aq^{-1}u^{-1}) }{u^2q^2-u^{-2}q^{-2}}\ ,\label{m1}\nonumber\\
\cW_-(u)&=& i \frac{\,q\,u\Phi_-( a\,u)\Phi_+(a q^{-1} u^{-1}) - q^{-1}u^{-1}\,\Phi_+( a\,u)\Phi_-(a q^{-1} u^{-1})}{u^2q^2-u^{-2}q^{-2}}\ ,\label{m2}\nonumber\\
\cG_+(u)&=& i k_-(q+q^{-1}) \,\Phi_+(a \,   u)\Phi_+(a q^{-1}u^{-1})-\frac{\rho}{(q-q^{-1})}\ ,\label{m3} \nonumber\\
\cG_-(u)&=& -i k_+(q+q^{-1})\Phi_-(a \,  u)\Phi_-(a q^{-1} u^{-1})-\frac{\rho}{(q-q^{-1})}\ \ . \label{m4}\nonumber
\een
\end{prop}

Infinite dimensional representations of the Zamolodchikov-Faddeev algebra (\ref{FZ}) have been studied in details (see for instance \cite{Luk} and references therein) in the context of quantum integrable systems. In particular, free field representations of Zamolodchikov-Faddeev operators are known at level $c=1$ \cite{JM,Pak}, given  a straightforward realization of $O_q(\widehat{sl_2})$ currents in terms of $q-$bosons.

\section{Comments}
Clearly, the isomorphisms (in the sense of coideal subalgebras) between the $q-$Onsager algebra, the $O_q(\widehat{sl_2})$ current algebra and the reflection equation algebra \cite{BasS} together with the present results suggest to consider in details the construction of vertex operators intertwining different representations of the $q-$Onsager algebra. Indeed, as will be shown in a forthcoming paper \cite{BasB2}, basic properties of such vertex operators are actually sufficient to derive the one-parameter family of solutions (\ref{Ku}) of the reflection equation algebra. Note that in this framework, the Zamolodchikov-Faddeev operators of \cite{Pak} provide one possible free field realization of the basic vertex operators. 
\vspace{1mm}

For the class of quantum integrable systems in the continuum or lattice which integrability condition is associated with the $q-$Onsager algebra,
these results open the possibility to develop an approach {\it \`a la Onsager} \cite{Bas0}  in cases in which the Hilbert space is infinite dimensional. Namely, inspired by previous works the idea is to solve the model solely using the $q-$Onsager algebra and its representation theory. This situation - which will be considered elsewhere - arises for instance while studying a solution of the semi-infinite XXZ open spin chain with {\it generic} boundary conditions: in the thermodynamic limit, the $q-$Onsager algebra emerges as the hidden non-Abelian symmetry of the Hamiltonian \cite{Bas2}. It is then not surprising that the one-parameter family of solutions (\ref{Ku}) of the reflection equation algebra here derived is nothing but the basic ingredient of the `renormalized' transfer matrix conjectured in \cite{JKKKM}. As a consequence, studying an approach {\it \`a la Onsager} for this model becomes essentially an extension to {\it generic} boundary conditions of the approach developed by Jimbo {\it et al.} \cite{JKKKM}. However, contrary to \cite{JKKKM} in which the analysis was restricted to the case of {\it diagonal} boundary conditions - the hidden non-Abelian symmetry of the Hamitonian being unidentified at that time -, the ground state structure in the asymptotic limit can be now studied according to the representation theory of the $q-$Onsager algebra.\vspace{1mm}
 
Finally, let us mention another interesting problem which concerns the extension of the results of Section 3 to the generalized $q-$Onsager algebras $O_q(\widehat{g})$ \cite{BasB}, that may bring some new insight in the study of affine Toda field theories or semi-infinite open spin chain (see e.g. \cite{FuKo,Ko}) with higher symmetries and scalar or dynamical boundary conditions \cite{BF}.   In this direction,  analogs of Miki's formula may be considered in order to derive free field realizations of $O_q(\widehat{g})$ algebras.  

\vspace{0.4cm}

\noindent{\bf Acknowledgements:}  
S.B. thanks LMPT for hospitality, where part of this work has been done, and INFN Iniziativa Specifica FI11 for financial support. S. B. thanks A. Molev, S. Pakuliak and E. Ragoucy for discussions. 
\vspace{1cm}

\begin{center} {\bf Appendix: Drinfeld-Jimbo presentation of $U_q(\widehat{sl_2})$ and R-matrix}\end{center}
\vspace{2mm}
Define the extended Cartan matrix $\{a_{ij}\}$ ($a_{ii}=2$,\ $a_{ij}=-2$ for $i\neq j$). The quantum affine algebra $U_{q}(\widehat{sl_2})$ is generated by the elements $\{H_j,E_j,F_j\}$, $j\in \{0,1\}$ which satisfy the defining relations
\beqa [H_i,H_j]=0\ , \quad [H_i,E_j]=a_{ij}E_j\ , \quad
[H_i,F_j]=-a_{ij}F_j\ ,\quad
[E_i,F_j]=\delta_{ij}\frac{q^{H_i}-q^{-H_i}}{q-q^{-1}}\
\nonumber\eeqa
together with the $q-$Serre relations
\beqa [E_i,[E_i,[E_i,E_j]_{q}]_{q^{-1}}]=0\ ,\quad \mbox{and}\quad
[F_i,[F_i,[F_i,F_j]_{q}]_{q^{-1}}]=0\ . \label{defUq}\eeqa
The sum $c=H_0+H_1$ is the central element of the algebra. The
Hopf algebra structure is ensured by the existence of a
comultiplication $\Delta: U_{q}(\widehat{sl_2})\mapsto U_{q}(\widehat{sl_2})\otimes U_{q}(\widehat{sl_2})$, antipode ${\cal S}: U_{q}(\widehat{sl_2})\mapsto U_{q}(\widehat{sl_2})$ 
and a counit ${\cal E}: U_{q}(\widehat{sl_2})\mapsto {\mathbb C}$ with
\beqa \Delta(E_i)&=&E_i\otimes q^{-H_i/2} +
q^{H_i/2}\otimes E_i\ ,\nonumber \\
 \Delta(F_i)&=&F_i\otimes q^{-H_i/2} + q^{H_i/2}\otimes F_i\ ,\nonumber\\
 \Delta(H_i)&=&H_i\otimes I\!\!I + I\!\!I \otimes H_i\ ,\label{coprod}
\eeqa
\beqa {\cal S}(E_i)=- q^{-1}E_i\ ,\quad {\cal S}(F_i)=-q F_i\ ,\quad {\cal S}(H_i)=-H_i \qquad {\cal S}({I\!\!I})=1\
\label{antipode}\nonumber\eeqa
and\vspace{-0.3cm}
\beqa {\cal E}(E_i)={\cal E}(F_i)={\cal
E}(H_i)=0\ ,\qquad {\cal E}({I\!\!I})=1\
.\label{counit}\nonumber\eeqa
Note that the opposite coproduct $\Delta'$ can be similarly defined with $\Delta'\equiv \sigma
\circ\Delta$ where the permutation map $\sigma(x\otimes y
)=y\otimes x$ for all $x,y\in U_{q}(\widehat{sl_2})$ is used.\vspace{2mm} 

The (evaluation) homomorphism  $\pi_u: U_{q}(\widehat{sl_2}) \mapsto \mathrm{End}({\cal V}_u)$ is chosen such that $({\cal V}\equiv{\mathbb C}^2)$
\beqa 
&&\pi_u[E_1]= uq^{1/2}\sigma_+\ , \qquad \ \ \ \ \ \pi_u[E_0]= uq^{1/2}\sigma_-\ , \nonumber\\
&&\pi_u[F_1]=
u^{-1}q^{-1/2}\sigma_-\ ,\qquad \pi_u[F_0]= u^{-1}q^{-1/2}\sigma_+\ ,\nonumber\\
\ &&\pi_u[q^{H_1}]= q^{\sigma_3}\ ,\qquad \qquad \quad  \ \pi_u[q^{H_0}]=
q^{-\sigma_3}\ \label{evalrep}
\eeqa
in terms of the Pauli matrices $\sigma_\pm,\sigma_3$:
\beqa
\sigma_+=\left(
\begin{array}{cc}
 0    & 1 \\
 0 & 0 
\end{array} \right) \ ,\qquad
\sigma_- =\left(
\begin{array}{cc}
 0    & 0 \\
 1 & 0 
\end{array} \right) \ ,\qquad
\sigma_3 =\left(
\begin{array}{cc}
 1    & 0 \\
 0 & -1 
\end{array} \right) \ .\label{Pauli}\nonumber
\eeqa

The $R-$matrix here considered is the solution of the intertwining equation:
\ben
R(u/v) (\pi_u \otimes \pi_v)\, \Delta(x)= (\pi_u \otimes \pi_v)\, \sigma \circ \Delta(x) R(u/v) \  .\nonumber
\een
According to above definitions and up to an overall scalar factor, it reads:
\begin{align}
R(u) =\left(
\begin{array}{cccc} 
1    & 0 & 0 & 0 \\
0  & \frac{u -  u^{-1}}{uq -  u^{-1}q^{-1}} & \frac{q-q^{-1}}{uq -  u^{-1}q^{-1}} & 0 \\
0  &  \frac{q-q^{-1}}{uq -  u^{-1}q^{-1}} &  \frac{u -  u^{-1}}{uq -  u^{-1}q^{-1}} &  0 \\
0 & 0 & 0 & 1
\end{array} \right) \ .\label{R}
\end{align}
In particular, we have
\beqa
R(u)R(1/u)=1\otimes 1 \ .\label{unit}
\eeqa
\vspace{1mm}

{\bf $\bullet$ Definition of $\tilde{R}(u)$}:
Specializing the universal R-matrix \cite{FR} to the tensor product of two finite dimensional representations, the $R-$matrix entering in the defining relations (\ref{RLL}) reads:
\beqa
\tilde{R}(u)=r(u)R(u)\  .\nonumber 
\eeqa
It satisfies the relation (\ref{unit}), together with the relation (see \cite{FR} for details): 
\ben
((((\tilde{R}(u))^{-1})^t)^{-1})^t&=&\tilde{R}(u q^{-2})\label{wcross} \ .\nonumber
\een
As a consequence, the scalar function $r(u)$ satisfies the functional relations:
\ben
r(u)r(u^{-1})=1, \quad r(u)&=&\frac{(u-u^{-1})(u q^{-2}-u^{-1}q^2)}{(u q^{-1}-u^{-1}q)^2}r(u q^{-2})\  .\label{refcr}
\een
\vspace{1mm}

{\bf $\bullet$ Definition of $\bar{R}(u)$}: For the Zamolodchikov-Faddeev algebra (\ref{FZ}), let us introduce the $R-$matrix
\beqa
\bar{R}(u)=f(u)R(u)\  .\nonumber 
\eeqa
It is assumed to satisfy (\ref{unit}) leading to $f(u)f(u^{-1})=1$, and the crossing symmetry relation:
\ben
\bar{R}(u)&=&M\otimes1\bar{R}^t (u^{-1}q^{-1})M\otimes1 \ \label{scross} \quad
\mbox{with}\quad M=\left(\begin{array}{cc} 0 & i\\ -i &0 \end{array}\right) \  .
\een
As a consequence, the scalar function satisfies the more restrictive relation (which implies (\ref{refcr})): 
\ben
f(u)=\frac{uq-u^{-1}q^{-1}}{u-u^{-1}}f(u^{-1}q^{-1})\ .\nonumber
\een

Note that using the functional relations for $r(u)$ and $f(u)$, defining 
\ben
\alpha(u) =\frac{(qu-q^{-1}u^{-1})(u q^{-1}-u^{-1}q)}{(u -u^{-1})^2}\nonumber
\een
one has:
\ben
\frac{r(u q^n)}{r(uq^{-n})}&=&\frac{f(u q^n)}{f(uq^{-n})}=\prod_{i=1}^{n}\alpha(uq^{n-2 i})\  \quad \mbox{with}  \quad n \in \mathbb Z_+ \ \label{ratios} \ .
\een

\end{document}